\documentclass[12pt]{article}

\usepackage{subcaption}
\usepackage{listings,jvlisting}
\usepackage{amsmath,amssymb}
\usepackage{bm}
\usepackage{graphicx, color}
\usepackage{wrapfig}
\usepackage{cite}
\usepackage{braket}
\usepackage{hyperref}
\usepackage{microtype}
\DeclareMathOperator{\Tr}{Tr}
\begin{document}
\title{
\begin{flushright}
\ \\*[-80pt]
\begin{minipage}{0.2\linewidth}
\normalsize
%arXiv:YYMM.NNNN \\
HUPD-2302 \\*[50pt]
\end{minipage}
\end{flushright}
{\Large \bf
$W$ Boson Mass and Grand Unification\\
via the Type-$\rm{I\hspace{-.01em}I}$ Seesaw-like Mechanism
}\\*[20pt]}

\author{
\centerline{
Yusuke~Shimizu $^{1,2}$\footnote{yu-shimizu@hiroshima-u.ac.jp}~and~Shonosuke~Takeshita $^{1}$\footnote{shonosuke@hiroshima-u.ac.jp}}
\\*[20pt]
\centerline{
\begin{minipage}{\linewidth}
\begin{center}
$^1${\it \normalsize
Physics Program, Graduate School of Advanced Science \\ and Engineering,~Hiroshima~University, \\
Higashi-Hiroshima~739-8526,~Japan \\*[5pt]
$^2${\it \normalsize
Core of Research for the Energetic Universe, Hiroshima University, \\
Higashi-Hiroshima 739-8526, Japan}
}
\end{center}
\end{minipage}}
\\*[50pt]}

\date{
\centerline{\small \bf Abstract}
\begin{minipage}{0.9\linewidth}
\medskip
\medskip
\small
We propose an SU(5) GUT model extended with two additional pairs of $\mathbf{10}$ representation vector-like fermions.
The CDF collaboration $W$ boson mass anomaly is explained by using the VEV of a real $\mathrm{SU(2)_L}$ triplet scalar coming from the $\mathbf{24}$ representation Higgs.
The vector-like fermions are decomposed partly into vector-like quark doublets.
Those vector-like quark doublets acquire mass from two sources;
through the Yukawa interaction with the real $\mathrm{SU(2)_L}$ triplet via a type-$\rm{I\hspace{-.01em}I}$ seesaw-like mechanism.
And, they acquire mass from the $\mathbf{24}$ representation Higgs.
% in addition to that of $\mathbf{24}$ representation Higgs.
We assume that the mass for the vector-like quark doublets is expressed in terms of the real triplets mass.
By combining the constraints on the vector-like quark masses with those on the heavy Higgs boson masses, we can obtain the narrow allowed mass ranges for the vector-like quark doublet and the real triplet.
Therefore, our model can be tested in searches for these particles in the near future.
In addition, the two additional pairs of vector-like fermions allow the SM gauge couplings to unify successfully at $M_{\mathrm{GUT}}\thickapprox5.1\times10^{15}$~GeV.
Our model is also testable by the future Hyper-Kamiokande experiment via the proton decay lifetime $\tau_p(p\to\pi^0{e^+})<1.0\times10^{35}$~years.
\end{minipage}
}

\begin{titlepage}
\maketitle
\thispagestyle{empty}
\end{titlepage}
\newpage
%------------------------------------------------------------------------------%
%--------------------------------   Introduction   --------------------------------%
%------------------------------------------------------------------------------%
\section{Introduction}
\label{sec:Intro}

Grand Unified Theory~(GUT) is an attractive theory to approach physics beyond the Standard Model~(SM).
GUT embeds the SM gauge groups $\mathrm{SU(3)_C\times{SU(2)_L}\times{U(1)_Y}}$ into a large simple group.
Therefore, GUT unifies the strong, weak, and electromagnetic forces.
In addition to the unification, GUT can explain the quantization of electric charge.
Furthermore, GUT predicts proton decay because the SM quarks and leptons are unified into irreducible representations of the GUT group.
Thus, GUT is testable by searching for proton decay.
The minimal GUT model is SU(5) one that was proposed by Georgi and Glashow in 1974~\cite{Georgi:1974sy}.
In the minimal SU(5) GUT model, the SM matter fields and gauge fields are unified into $\mathbf{\bar{5}}$, $\mathbf{10}$, and $\mathbf{24}$ adjoint representations.
Also, the scalar fields of this model are in the $\mathbf{5}$ and $\mathbf{24}$ representations.
These fields realize spontaneous symmetry breaking by acquiring vacuum expectation values~(VEVs).
However, the minimal SU(5) GUT model is ruled out on the following two reasons:
\begin{itemize}
  \item First, it fails to unify the SM gauge couplings successfully at high energy by only the SM fields.
  \item Second, the proton lifetime predicted by the minimal SU(5) GUT model is $\tau_p(p\to\pi^{0}e^+)\approx10^{30}$-$10^{31}$~years~\cite{Georgi:1974yf}, which is inconsistent with the current experimental result reported by Super-Kamiokande experiment of $\tau_p(p\to\pi^{0}e^+)>2.4\times10^{34}$~years~\cite{Super-Kamiokande:2020wjk}.
\end{itemize}

Last year, the CDF collaboration reported a new result of the $W$ boson mass, $M_W^{\mathrm{CDF}}=80.4335\pm0.0094$~GeV~\cite{CDF:2022hxs}.
The average of the new CDF results combined with the previous results from LEP2, Tevatron, LHC, and LHCb experiments is $M_W^{\mathrm{exp}}=80.4133\pm0.0080$~GeV~\cite{deBlas:2022hdk}.
This combined value is 6.5 $\sigma$ deviations away from the SM prediction of $M_W^{\mathrm{SM}}=80.3500\pm0.0056$~GeV~\cite{deBlas:2022hdk}.
Hence, physics beyond the SM is required to explain this shift of the $W$ boson mass.
A candidate for explaining the $W$ boson mass anomaly is adding a real $\mathrm{SU(2)_L}$ triplet scalar with zero hypercharge~\cite{Ross:1975fq, Gunion:1989ci, Lynn:1990zk, Blank:1997qa, Forshaw:2003kh, Chen:2006pb, Chankowski:2006hs, Chivukula:2007koj, Bandyopadhyay:2020otm, FileviezPerez:2022lxp, Wu:2022uwk}.
The $\mathrm{SU(2)_L}$ triplet vacuum expection value~(VEV) contributes to the $W$ boson mass at tree level.
In the minimal SU(5) GUT model, the real $\mathbf{24}$ representation Higgs contains a real $\mathrm{SU(2)_L}$ triplet scalar with zero hypercharge.
Therefore, the SU(5) GUT can explain the $W$ boson mass anomaly by using the VEV of a real triplet in the $\mathbf{24}$ representation Higgs~\cite{Evans:2022dgq, Senjanovic:2022zwy, Calibbi:2022wko}.
Some other beyond the SM candidates for explaining the $W$ boson mass anomaly are built in the framework of SO(10) GUT~\cite{Fritzsch:1974nn} or the type-$\rm{I\hspace{-.01em}I}$ seesaw mechanism~\cite{Magg:1980ut, Cheng:1980qt, Lazarides:1980nt, Mohapatra:1980yp} are shown in Refs.~\cite{Lazarides:2022spe, Chao:2022blc}.

%In this paper, we propose an SU(5) GUT model added two pairs of $\mathbf{10}$ representation for the vector-like fermions to the minimal SU(5) GUT model.
In this paper, we propose an SU(5) GUT model extended with two additional pairs of $\mathbf{10}$ representation vector-like fermions\footnote{In Ref.~\cite{Evans:2022dgq}, authors mentioned that the SU(5) GUT model added one pair of $\mathbf{10}$ representation for the vector-like fermions to the minimal SU(5) GUT model is the minimal one.
In our work, we can get the narrow allowed mass ranges for relevant particles.
Under these conditions, the SM gauge couplings do not unify successfully.
Therefore, we add two pairs of $\mathbf{10}$ representation for the vector-like fermions to the minimal SU(5) GUT model.}.
%By a real $\mathrm{SU(2)_L}$ triplet coming from the $\mathbf{24}$ representation Higgs getting the VEVs, the $W$ boson mass anomaly reported by the CDF collaboration can be explained.
The CDF collaboration $W$ boson mass anomaly is explained by using the VEV of a real $\mathrm{SU(2)_L}$ triplet scalar coming from the $\mathbf{24}$ representation Higgs.
%We assume that the mass for the vector-like quark doublet is expressed in terms of a real triplet mass
The vector-like fermions are decomposed partly into vector-like quark doublets.
%The vector-like quark doublet acquire the mass through the Yukawa interaction of a real triplet via the type-$\rm{I\hspace{-.01em}I}$ seesaw-like mechanism in addition to that of $\mathbf{24}$ representation Higgs.
Those vector-like quark doublets acquire mass from two sources;
through the Yukawa interaction with the real $\mathrm{SU(2)_L}$ triplet via a type-$\rm{I\hspace{-.01em}I}$ seesaw-like mechanism.
And, they acquire mass from the $\mathbf{24}$ representation Higgs.
We assume that the mass for the vector-like quark doublets is expressed in terms of the real triplets mass through the Yukawa interaction of a real triplet.
Because a real triplet and heavy Higgs boson masses are almost the same, we can get the constraints about the vector-like quark doublet mass by considering the constraints of the heavy Higgs bosons.
%As a result of combining the constraints of the vector-like quark doublet mass with that of the heavy Higgs boson masses, we can obtain the allowed mass ranges for the vector-like quark doublet and a real triplet scalar which are very narrow.
By combining the constraints on the vector-like quark masses with those on the heavy Higgs boson masses, we can obtain the narrow allowed mass ranges for the vector-like quark doublet and the real triplet.
%Therefore, our model can be tested by the search of these particles in the near future.
Therefore, our model can be tested in searches for these particles in the near future.

We set a benchmark on the mass eigenvalues for the relevant particles and solve the renormalization group equations~(RGE).
The new contributions from the two additional pairs of vector-like fermions allow the SM gauge couplings to unify successfully at $M_{\mathrm{GUT}}\thickapprox5.1\times10^{15}$~GeV.
%In the case that included the new contributions, the SM gauge couplings unify successfully at $M_{\mathrm{GUT}}\thickapprox5.1\times10^{15}$~GeV and
At this benchmark, the value for the unified gauge couplings is $\alpha_{\mathrm{GUT}}=\alpha_1=\alpha_2=\alpha_3\thickapprox1/34.7$.
In addition, the SM Higgs quartic coupling is positive for all energy scales.
Hence, the SM Higgs potential is stabilized.
Our model also predicts the proton lifetime as $\tau_p(p\to\pi^0{e^+})\approx{4.12\times10^{34}}$~years, for proton decay mediated by the SU(5) gauge bosons.
This is testable by future proton decay searches, for example the Hyper-Kamiokande experiment expects an upper limit of proton lifetime as $\tau_p(p\to\pi^0{e^+})<1.0\times10^{35}$~years~\cite{Dealtry:2019ldr}.

This paper is organized as follows.
In section~\ref{sec:Wmass}, we discuss the model that contains a real $\mathrm{SU(2)_L}$ triplet scalar with zero hypercharge, and derive the mass eigenvalues for the physical scalars.
We also obtain the real triplet VEV to explain the $W$ boson mass anomaly.
In section~\ref{sec:themodel}, we propose an extension to the minimal SU(5) GUT model and consider the allowed mass ranges of the new particles.
Section~\ref{sec:protondecay} is shown our results for gauge unification and proton lifetime.
We also discuss the testability of our model.
Section~\ref{sec:Summary} is devoted to summary.
In appendix~\ref{sec:beta_coefficients}, we show the RGE formulae and the beta coefficients of the relevant particles.

%------------------------------------------------------------------------------%
%---------------------------------The W boson mass----------------------------------------%
%------------------------------------------------------------------------------%
\section{The $W$ boson mass}
\label{sec:Wmass}

In this section, we discuss the scalar sector with the SM Higgs $H$ and a real $\mathrm{SU(2)_L}$ triplet $T$ with zero hypercharge.
The $H$ and $T$ are given by
\begin{equation}
\label{eq:Scalar}
    H=
    \begin{pmatrix}
       \phi^{+}\\
       \phi^0\\
    \end{pmatrix},\quad
    T=\frac{1}{2}
    \begin{pmatrix}
        T^0&\sqrt{2}T^{+}\\
        \sqrt{2}T^{-}&T^0\\
    \end{pmatrix}.
\end{equation}
The Lagrangian for the scalar sector is 
\begin{equation}
\label{eq:scalarlagrangian}
\mathcal{L}_{\mathrm{scalar}}\supset(D_\mu{H})^{\dagger}(D_\mu{H})+\Tr(D_\mu{T})^{\dagger}(D_\mu{T})-V(\Phi,T),
\end{equation}
where
\begin{equation}
\label{eq:tripletcovariant}
    D_\mu{T}=\partial_\mu{T}+\mathrm{i}g_2[W_\mu,T].
\end{equation}
Here, $W_\mu$ and $g_2$ are the $\mathrm{SU(2)_L}$ gauge bosons and coupling.
The most general scalar potential is given by
\begin{equation}
\begin{aligned}
\label{eq:scalarpotential}
    V(H,T)=&-m_h^2{H}^\dagger{H}+\lambda_{0}({H}^\dagger{H})^2+M_T^2\Tr{T}^2+\lambda_1\Tr{T}^4+\lambda_2(\Tr{T}^2)^2\\
&+\alpha(H^\dagger{H})\Tr{T}^2+\beta{H}^\dagger{T^2}{H}+\mu{H}^\dagger{T}{H}.
\end{aligned}    
\end{equation}
The $H$ and $T$ acquire the VEVs at the minimum of the potential.
Then, we can parametrize the scalars as,
\begin{equation}
\label{eq:scalarparameter}
    H=
    \begin{pmatrix}
        \phi^{+}\\
        (v_h+h^0+\mathrm{i}G^0)/\sqrt{2}\\
    \end{pmatrix},\quad
    T=\frac{1}{2}
    \begin{pmatrix}
        v_T+t^0&\sqrt{2}t^{+}\\
        \sqrt{2}t^{-}&-v_T-t^0\\
    \end{pmatrix},
\end{equation}
where $v_h$ and $v_T$ are the VEV of the scalar fields, $H$ and $T$.
By using Eqs.~\eqref{eq:scalarpotential} and \eqref{eq:scalarparameter}, the minimization conditions for the scalar potential are written as
\begin{align}
    \label{eq:minimizationh}
    m_h^2&=\lambda_0{v_h^2}+\frac{A}{2}v_h^2-\frac{\mu}{2}v_T,\\
    \label{eq:minimizationT}
        M_T^2&=\frac{\mu{v_h^2}}{4v_T}-\frac{A}{2}v_h^2-\frac{B}{2}v_T^2,
\end{align}
where 
\begin{equation}
    A=\alpha+\frac{\beta}{2},\quad{B}=\lambda_1+2\lambda_2.
\end{equation}
The real scalar mass matrix in the $(h^0,t^0)$ basis is 
\begin{equation}
\label{eq:realscalarmatrix}
    \mathcal{M}_0^2=
    \begin{pmatrix}
        2\lambda_0{v_h^2}&{A}v_h{v_T}-\frac{\mu{v_h}}{2}\\
        {A}v_h{v_T}-\frac{\mu{v_h}}{2}&Bv_T^2+\frac{\mu{v_h^2}}{4v_T}
    \end{pmatrix}.
\end{equation}
And the charged scalar mass matrix in the $(\phi^{\pm},T^{\pm})$ basis is
\begin{equation}
    \mathcal{M}_{\pm}^2=
    \begin{pmatrix}
        \mu{v_T}&\frac{\mu{v_T}}{2}\\
        \frac{\mu{v_T}}{2}&\frac{\mu{v_h^2}}{4v_T}\\
    \end{pmatrix}.
\end{equation}
The mass eigenstates are written in terms of the gauge eigenstates as
\begin{align}
    \begin{pmatrix}
        h\\
        H\\
    \end{pmatrix}&=
    \begin{pmatrix}
        \cos{\theta_0}&\sin{\theta_0}\\
        -\sin{\theta_0}&\cos{\theta_0}\\
    \end{pmatrix}
    \begin{pmatrix}
        h^0\\
        t^0\\
    \end{pmatrix},\\
    \begin{pmatrix}
        H^{\pm}\\
        G^{\pm}\\
    \end{pmatrix}&=
    \begin{pmatrix}
        -\sin{\theta_+}&\cos{\theta_+}\\
        \cos{\theta_+}&\sin{\theta_+}\\
    \end{pmatrix}
    \begin{pmatrix}
        \phi^{\pm}\\
        T^{\pm}\\
    \end{pmatrix},
\end{align}
where the mixing angles between the mass and gauge eigenstates are 
\begin{align}
\label{eq:mixinganglereal}
    \tan{2\theta_0}&=\frac{4{v_h}v_T(-\mu+2A{v_T})}{8\lambda_0{v_h^2}v_T-4B{v_T^3}-\mu{v_h^2}},\\
\label{eq:mixinganglecharged}    
    \tan{2\theta_+}&=\frac{4{v_h}v_T}{4{v_T^2}-v_h^2}.
\end{align}
In the limit $v_h\gg{v_T}$, we get $\theta_0\ll{1}$ in Eq.~\eqref{eq:mixinganglereal}.
Then, the mass eigenvalues for the physical scalars are
\begin{align}
\label{eq:SMHiggs}
    M_h^2&=2\lambda_0{v_h^2},\\
\label{eq:HeavyHiggs}    
    M_H^2&=B{v_T^2}+\frac{\mu{v_h^2}}{4{v_T}},\\
\label{eq:ChargedHiggs}    
    M_{H^\pm}^2&=\mu{v_T}\left(1+\frac{v_h^2}{4{v_T^2}}\right).
\end{align}
In Eqs.~\eqref{eq:SMHiggs}-\eqref{eq:ChargedHiggs}, $h$ is the SM-like Higgs.
We find that in the limit $v_h\gg{v_T}$, the masses of the scalars are $M_{H^\pm}={M_H}={M_T}\approx\mu{v_h^2}/4{v_T}$.

The VEV of the real $\mathrm{SU(2)_L}$ triplet scalar contributes to the $W$ boson mass.
At tree level, the $W$ boson mass is given by
\begin{equation}
\label{eq:wbosoneq}
    M_W^2=(M_W^{\mathrm{SM}})^2+g_2^2{v_T^2},
\end{equation}
where $M_W^{\mathrm{SM}}$ is the $W$ boson mass in the SM.
The VEV of the real $\mathrm{SU(2)_L}$ triplet scalar does not contribute to the $Z$ boson mass.
Then, we assume the SM expectation value for the $Z$ boson mass.
The average of the $W$ boson mass obtained by combining the CDF results with previous results from LEP2, Tevatron, LHC, and LHCb experiments~\cite{deBlas:2022hdk} is
\begin{equation}
    M_W^{\mathrm{exp}}=80.4133\pm{0.0080}~\text{GeV}.
\end{equation}
This value deviates from the SM prediction~\cite{deBlas:2022hdk} by $6.5~\sigma$,
\begin{equation}
    M_W^{\mathrm{SM}}=80.3500{\pm}0.0056~\text{GeV}.
\end{equation}
In order to explain the deviation with Eq.~\eqref{eq:wbosoneq}, we use $g_2(\mu=M_Z)=0.657452$~\cite{deBlas:2022hdk, ParticleDataGroup:2022pth}, and the $v_T$ must be
\begin{equation}
    \label{eq:realtripletvev}
    v_T=4.85~\text{GeV}.
\end{equation}
Then, by using Eq.~\eqref{eq:mixinganglecharged}, we can get the mixing angle for the charged scalar sector,
\begin{equation}
    \theta_{+}=-0.0394.
\end{equation}
The couplings to the charged Higgs and its branching ratios are fixed with $v_T$ and $\theta_+$~\cite{FileviezPerez:2022lxp}.

%------------------------------------------------------------------------------%
%-------------------------------Model-------------------------------%
%------------------------------------------------------------------------------%
\section{The SU(5) GUT model}
\label{sec:themodel}
In previous section, we discussed the $W$ boson mass anomaly with the SM Higgs and a real $\mathrm{SU(2)_L}$ triplet scalar with zero hypercharge.
In this section, we build an SU(5) GUT model to explain the $W$ boson mass anomaly.
The field contents of the minimal SU(5) GUT model~\cite{Georgi:1974sy} are $\bar{5}_L^{i}$ and $10_L^{i}$ for the SM matter fields, $5_H$ and $24_H$ for the scalar sector, and $A_\mu$ for the gauge field, respectively.
The $24_H$ is composed of,
\begin{equation}
24_{H}\sim\Phi(8,1,0)\oplus\Phi(1,3,0)\oplus\Phi(1,1,0)\oplus\Phi(3,2,-\frac{5}{6})\oplus\Phi(\bar{3},2,\frac{5}{6}).\\
\end{equation}
The $24_H$ includes a real $\mathrm{SU(2)_L}$ triplet scalar with zero hypercharge.
Thus, the minimal SU(5) GUT model can explain the $W$ boson mass anomaly via that real triplet as discussed in section \ref{sec:Wmass}.
However, this model has problems.
Even if the real triplet exists at low energy, and contributes to the RGE, the SM gauge couplings do not unify successfully.
Additionally, this minimal model is excluded by the proton decay search~\cite{Evans:2022dgq}.
Therefore, this model needs an extension.

We propose extending the minimal SU(5) GUT model with two additional pairs of $\mathbf{10}$ representation vector-like fermions.
The reasons for including two additional pairs are discussed in section \ref{sec:protondecay}.
We denote the vector-like fermions as $10_{L,R}^{4,5}$ and the decompositions of these are
\begin{equation}
    \label{eq:vectorlikedecomp}
    10_{L,R}^{4,5}=Q(3,2,1/6)\oplus{U^c}(\Bar{3},1,-2/3)\oplus{E^c}(1,1,1).
\end{equation}
We impose a $\mathbb{Z}_3$ symmetry to forbid mixing between the SM, 4th vector-like, and 5th vector-like fermions.
Then, we can write the Yukawa interactions for the SM and vector-like fermions sector as
\begin{align}
\label{eq:SM_Lagrangian}
\mathcal{L}_{\mathrm{SM}}\supset\sum_{i,j=1}^{3}Y^{ij}_{1}5_H{10}_L^i{10_L^j}+\sum_{i,j=1}^{3}Y^{ij}_{2}5_H^*\bar{5}_L^i{10}_L^j+\mathrm{h.c.},\\
\label{eq:VL_Lagrangian}
\mathcal{L}_{\mathrm{VL}}\supset\overline{10}_L^4[{Y}_{10}^{4}{24}_H+M^4_{10}]10_{R}^4+\overline{10}_L^5[{Y}_{10}^5{24}_H+M_{10}^5]10_{R}^5+\mathrm{h.c.},
\end{align}
where $Y_{1,2}^{ij}$ are the Yukawa couplings for the SM sector, $Y_{10}^{4,5}$ are the Yukawa couplings for the vector-like fermions sector, and $M_{10}^{4,5}$ are the masses for 4th and 5th vector-like fermions, respectively.
In the minimal SU(5) GUT model, the mass relations among the down-type quarks and charged leptons are inconsistent with the experimental results.
The simple way to correct the mass relations is to add non-renormalizable terms to the Lagrangian~\cite{Ellis:1979fg, Dorsner:2005fq, Dorsner:2006hw}.
An example term is $Y_u^{ij}24_H{5_H}10^i_L{10^j_L}/\Lambda$, where $\Lambda$ is the cut-off scale.
Another way to correct the relations, is to add the $\mathbf{45}$ representation Higgs~\cite{Georgi:1979df, Kalyniak:1982pt, Eckert:1983bn}.
Moreover, the scalars composed of the $\mathbf{45}$ representation Higgs help to achieve gauge unification~\cite{FileviezPerez:2007bcw, Dorsner:2007fy, FileviezPerez:2016sal, Boucenna:2017fna, FileviezPerez:2018dyf, FileviezPerez:2019fku, FileviezPerez:2019ssf, Shimizu:2022wsk}.

In our model, the VEV of the real triplet from the $24_H$ can be used to explain the $W$ boson mass anomaly.
This usage is the same as section \ref{sec:Wmass}.
Then, vector-like quark doublet of Eq.~\eqref{eq:vectorlikedecomp} acquires mass through the Yukawa interaction of the real triplet via the type-$\rm{I\hspace{-.01em}I}$ seesaw-like mechanism.
The Lagrangian for this interaction is
\begin{equation}
\label{eq:VLRTYukawa}
    \mathcal{L}_{\mathrm{VL-RT}}\supset{Y_{10}^4}T\Bar{Q}^4_{L}{Q^4_R}+{Y_{10}^5}T\Bar{Q}^5_{L}{Q^5_R}+\mathrm{h.c.},
\end{equation}
where $T$ is the real triplet from the $24_H$, and $Q_{L,R}^{4,5}$ is vector-like quark doublet in the $10_{L,R}^{4,5}$.
Spontaneous symmetry breaking occurs by the following steps:
\begin{enumerate}
  \item $24_H$ breaks the SU(5) symmetry with the VEVs as $\langle24_H\rangle=V/2\sqrt{15}~\mathrm{Diag}(-2,-2,-2,3,3)$.
  \item $5_H$ realizes electroweak symmetry breaking by taking the VEVs as $\langle{5_H}\rangle=(0,0,0,0,v_h/\sqrt{2})$.
  \item By using Eq.~\eqref{eq:minimizationT}, we derive the VEV of the real triplet, $v_T\approx\mu{v_h^2}/4M_T^2$.
\end{enumerate}
After spontaneous symmetry breaking, we can acquire the mass eigenvalues for the vector-like fermions as follows:
\begin{align}
\label{eq:quarkdoubletVmass4}
    M_Q^4&=M^4_{10}-\frac{Y^4_{10}}{4\sqrt{15}}V+Y^4_{10}\frac{\mu{v_h^2}}{8M_T^2},\\
\label{eq:upquarkVmass4}    
    M_U^4&=M^4_{10}+\frac{Y^4_{10}}{\sqrt{15}}V,\\
\label{eq:electronVmass4}    
    M_E^4&=M^4_{10}-\frac{3Y^4_{10}}{2\sqrt{15}}V,\\
\label{eq:quarkdoubletVmass5}
    M_Q^5&=M^5_{10}-\frac{Y^5_{10}}{4\sqrt{15}}V+Y^5_{10}\frac{\mu{v_h^2}}{8M_T^2},\\
\label{eq:upquarkVmass5}    
    M_U^5&=M^5_{10}+\frac{Y^5_{10}}{\sqrt{15}}V,\\
\label{eq:electronVmass5}    
    M_E^5&=M^5_{10}-\frac{3Y^5_{10}}{2\sqrt{15}}V.
\end{align}
Here, $M^4_{10}$ and $Y^4_{10}$ are free parameters.
Then, assuming that $M^4_{10}=Y^4_{10}V/(4\sqrt{15})$, we can express the mass eigenvalue for the 4th vector-like quark doublet as $M^4_Q=Y^4_{10}{\mu}{v_h^2}/(8M_T^2)$.
A consequence is the 4th vector-like right-handed up quark and electron have the same mass as $M_U=M_E=5Y^4_{10}V/(4\sqrt{15})$.

In Eq.~\eqref{eq:quarkdoubletVmass4}, the 4th vector-like quark doublet mass $M_Q$ is expressed in terms of the real triplet mass $M_T$.
In section \ref{sec:Wmass}, we found $M_{H^\pm}={M_H}={M_T}\approx\mu{v_h^2}/4{v_T}$.
Thus, we can obtain constraints on the mass of the 4th vector-like quark doublet by considering constraints on the heavy Higgs bosons.
We can get an upper bound for the mass of the heavy Higgs bosons from the perturbative unitarity of the $WW$ scattering cross-section~\cite{Chivukula:2007koj},
\begin{equation}
\label{eq:Higgsupper}
    M_{H,H^{\pm}}\lesssim\frac{2\sqrt{\pi}v_h^2}{v_T}.
\end{equation}
In order to explain the $W$ boson mass, we fixed the VEV for the real triplet as $v_T=4.85$~GeV in Eq.~\eqref{eq:realtripletvev}.
By combining that information with Eq.~\eqref{eq:Higgsupper}, we get an upper bound for the mass of heavy Higgs bosons, $M_{H,H^{\pm}}\lesssim~44.3$~TeV.
In Eqs.~\eqref{eq:HeavyHiggs} and \eqref{eq:ChargedHiggs}, we derive the mass of the heavy Higgs bosons as $M_H^2={M_{H^{\pm}}^2}\approx\mu{v_h^2}/4{v_T}$.
As a result, the upper bound of the trilinear coupling $\mu$ is given by
\begin{equation}
    \label{eq:trilinearcouplingconstraint}
    \mu<\frac{16\pi{v_h^2}}{v_T}\approx{6.28\times{10}^2}~\text{TeV}.
\end{equation}
In our assumption of $M^4_{10}=Y^4_{10}V/(4\sqrt{15})$, the mass of the 4th vector-like quark doublet from Eq.~\eqref{eq:quarkdoubletVmass4} is $M^4_Q\approx{{Y^4_{10}}\mu{v_h^2}}/(8M_T^2)$.
Having constrained the trilinear coupling in Eq.~\eqref{eq:trilinearcouplingconstraint}, we can derive the upper bound among the 4th vector-like quark doublet and a real $\mathrm{SU(2)_L}$ triplet scalar mass as
\begin{equation}
\label{eq:massupperbound}
    M^4_{Q}\times{(M_T)^2}<4.76\times{Y^4_{10}}~\text{(TeV)}^3.
\end{equation}
Thus, the masses of the new particles are restricted from this upper bound.

Next, we consider the experimental constraints for the new particles.
The lower bound of the mass for vector-like quarks is 1660~GeV~\cite{CMS:2018dcw}.
We also get a bound on the heavy neutral Higgs boson mass of $M_{H}>1400$~GeV~\cite{ATLAS:2019tpq}.
In addition, we need to consider the bound of the charged Higgs boson mass as $M_{H^+}>1000$~GeV~\cite{ATLAS:2021upq}~\footnote{In our model, the real triplet mass and heavy neutral and charged Higgs bosons have approximately the same mass.}.
By combining the upper bound for the vector-like quark doublet and the real triplet masses in Eq.~\eqref{eq:massupperbound} with the bounds for each particle, we obtain the allowed mass ranges.
However, the upper bound in Eq.~\eqref{eq:massupperbound} depends on the Yukawa coupling $Y^4_{10}$.
This is important because, if $Y^4_{10}$ is small, the upper bound is inconsistent with the experimental constraints.
To avoid that inconsistency, we assume $Y^4_{10}$ is exactly one.

Based on the above, we consider the upper bound on the mass for the real $\mathrm{SU(2)_L}$ triplet scalar.
In order to obtain this upper bound, we combine the upper bound in Eq.~\eqref{eq:massupperbound} with the bound for the vector-like quark mass.
Then, the upper bound of the mass for the real $\mathrm{SU(2)_L}$ triplet scalar is $M_{T}<1693$~GeV.
On the other hand, by combining the upper bound in Eq.~\eqref{eq:massupperbound} with the bound for the heavy neutral or charged Higgs boson masses, we can obtain the upper bound of the mass for the 4th vector-like quark doublet as $M^4_{Q}<2428$~GeV~(the heavy neutral Higgs boson limit) or $M^4_{Q}<4759$~GeV~(the charged Higgs boson limit).
As the result, we can get the allowed mass range for the 4th vector-like quark doublet and a real triplet scalar as
\begin{align}
  \label{eq:heavyneutralHiggslimit}
    1660<M^4_Q<2428\text{ GeV,} && 1400<M_T<1693\text{ GeV,} && \text{heavy neutral Higgs boson limit}
    \\
  \label{eq:chargedHiggslimit}
    1660<M^4_Q<4759\text{ GeV,} && 1000<M_T<1693\text{ GeV,} && \text{charged Higgs boson limit}
\end{align}
In the next section, we use these mass ranges for the discussion of the proton decay and gauge unification.

%------------------------------------------------------------------------------%
%-------------------------Proton decay and gauge unification-----------------------%
%------------------------------------------------------------------------------%
\section{Proton decay and gauge unification}
\label{sec:protondecay}

In this section, we evaluate the contribution of the new particles to the RGE and estimate the proton lifetime.
In the previous section, we obtained the masses for the vector-like fermion from Eq.~\eqref{eq:quarkdoubletVmass4} to Eq.~\eqref{eq:electronVmass5}.
By assuming that $M^4_{10}=Y^4_{10}V/(4\sqrt{15})$, the 4th vector-like quark doublet mass is written as $M^4_Q=Y^4_{10}{\mu}{v_h^2}/(8M_T^2)$.
A consequence of that assumption is the 4th vector-like right-handed up quark and electron have the same mass $M^4_U=M^4_E=5Y^4_{10}V/(4\sqrt{15})$.
In addition, we obtained the allowed mass ranges for the 4th vector-like quark doublet and the real triplet in Eq.~\eqref{eq:massupperbound}.
In the instance that $Y^4_{10}$ is exactly one, the allowed mass range for the 4th vector-like quark doublet and a real triplet scalar are given in Eqs.~\eqref{eq:heavyneutralHiggslimit} and \eqref{eq:chargedHiggslimit}.

In the models with one additional pair for the $\mathbf{10}$ representation vector-like fermions, there are five candidates for contributing to the RGEs;
the 4th vector-like fermions, $Q^4$, $U^4$, and $E^4$, the real triplet scalar $T$ and the color octet scalar ${H_8}$ in the $24_H$.
We assumed that $Y^4_{10}$ is exactly one in order not to inconsistent with the experimental constraints.
In that case, the 4th vector-like quark doublet and a real triplet scalar masses are restricted by the narrow allowed mass ranges in Eqs.~\eqref{eq:heavyneutralHiggslimit} and \eqref{eq:chargedHiggslimit}.
In addition, the 4th vector-like right-handed up quark and  electron have the same GUT scale mass.
Under these conditions, the SM gauge couplings do not unify successfully at high energy.
For the above reason, we include two additional pairs of $\mathbf{10}$ representation vector-like fermions to the minimal SU(5) GUT model.

The decay width for a proton decaying into a charged lepton mediated by SU(5) gauge bosons is given by~\cite{Nath:2006ut, FileviezPerez:2016sal, FileviezPerez:2018dyf, Calibbi:2022wko};
\begin{align}
\label{eq:protonwidth}
     \Gamma({p\rightarrow{\pi^0{e^+_\beta}}})
=\frac{\pi{m_p}\alpha^2_{\mathrm{GUT}}}{2M^4_{\mathrm{GUT}}}{A^2}\Big[|c(e,d^c)\langle\pi^0|(ud)_R{u_L}|{p}\rangle|^2+|c(e^c,d)\langle\pi^0|(ud)_L{u_L}|p\rangle|^2\Big],
\end{align}
where
\begin{equation}
A=A_{\mathrm{QCD}}A_{\mathrm{SR}}=\left(\frac{\alpha_3(m_b)}{\alpha_3(M_Z)}\right)^{6/23}\left(\frac{\alpha_3(Q)}{\alpha_3(m_b)}\right)^{6/25}\left(\frac{\alpha_3(M_Z)}{\alpha_3(M_{\mathrm{GUT}})}\right)^{2/7}.
\end{equation}
In Eq.~\eqref{eq:protonwidth}, $m_p=0.938$~GeV is the proton mass~\cite{ParticleDataGroup:2022pth}, $\alpha_{\mathrm{GUT}}$ is the value for the unified gauge couplings at high energy $M_{\mathrm{GUT}}$, and $A$ defines the running of the operators.
$A_{\mathrm{QCD}}$ denotes the running from the $M_Z$ to the scale $Q\approx2.3$~GeV and $A_{\mathrm{SR}}$ the running from $M_\mathrm{GUT}$ to the electroweak scale.
We use the numerical values of $A_{\mathrm{QCD}}\approx{1.2}$ and $A_{\mathrm{SR}}\approx{1.5}$~\cite{Nath:2006ut}.
The coefficients in Eq.~\eqref{eq:protonwidth} are given by~\cite{Nath:2006ut, FileviezPerez:2004hn}
\begin{align}
    c(e^c_\alpha,d_\beta)&=V_1^{11}V_2^{\alpha\beta}+(V_1{V_{UD}})^{1\beta}(V_2{V^\dagger_{UD}})^{\alpha1},\\
    c(e_\alpha,d^c_{\beta})&=V_1^{11}V_3^{\beta\alpha}.
\end{align}
The $V$'s are the mixing matrices defined as
\begin{equation}
V_1=U^\dagger_{C}U,\quad{V_2}=E^\dagger_{C}D,\quad{V_3}=D^\dagger_{C}E,\quad{V_{UD}}=U^\dagger{D},
\end{equation}
where the matrices $U, E,$ and $D$ diagonalize the Yukawa couplings
\begin{equation}
U_{C}^{T}Y_u{U}=Y_u^{\mathrm{Diag}},\quad{D_{C}^{T}Y_d{D}=Y_d^{\mathrm{Diag}}},\quad{E_{C}^{T}Y_e{E}=Y_e^{\mathrm{Diag}}}.
\end{equation}
The $\langle\pi^0|(ud)_L{u_R}|p\rangle$ and $\langle\pi^0|(ud)_R{u_L}|{p}\rangle$ are the hadronic matrix elements computed in lattice QCD calculations and the numerical values are~\cite{Aoki:2017puj}
\begin{equation}
    \langle\pi^0|(ud)_L{u_L}|p\rangle=0.134(5)(16)~\mathrm{GeV}^2,\quad\langle\pi^0|(ud)_R{u_L}|{p}\rangle=-0.131(4)(13)~\mathrm{GeV}^2.
\end{equation}
Note that we assume the correct mass relation between down-type quarks and charged leptons by introducing non-renormalizable operators.
Thus, in our model, the proton lifetime depends on unknown mixing, $V_2$ and $V_3$ and it is possible that the flavor structure in the fermionic mass matrices suppress the proton decay rate~\cite{Nath:2006ut, Dorsner:2004xa}.
We use the numerical values as $c(e^c_\alpha,d_\beta)=1+|V_{UD}|^2\simeq1.95$ where $V_{UD}$ is the $(1,1)$ element of $V_{\mathrm{CKM}}$, and $c(e_\alpha,d^c_{\beta})=1$.
%In our analysis of solving the RGE, we fix the masses such that the 4th vector-like quark doublet have mass $M^4_Q=2000$~GeV, the 5th vector-like right-handed up quark have mass $M^5_U=5.0\times{10}^{8}$~GeV, the real triplet scalar and color octet scalar have masses $M_T=M_{H_8}=1500$~GeV, and the other particles have tha GUT scale masses, respectively.

In our model, the candidates for contributing to the RGE are the vector-like fermions $Q^{4,5}$, $U^5$, and $E^5$, the real triplet scalar $T$ and the color octet scalar $H_8$ in the $24_H$.
We assume the color triplet scalar field contained in $5_H$ has a GUT scale mass because this field can mediate proton decay.
In our analysis of proton lifetime and gauge unification, we consider the 4th vector-like quark doublet and the real triplet scalar masses in the allowed mass range of Eq.~\eqref{eq:heavyneutralHiggslimit}.
This means we freely vary the masses of $Q^5$, $U^5$, $E^5$, and $H_8$.
%\footnote{The 5th vector-like quark doublet and right-handed electron masses are expressed in terms of the 5th vector-like right-handed up quark mass $M_U^5$ and Yukawa couplings $Y^5_{10}$.}
In addition, we evaluate the contributions of the SM particles to the RGEs at the 2-loop level~\cite{Machacek:1983tz, Machacek:1983fi, Machacek:1984zw} and the new particles contributions at the 1-loop level.
See appendix~\ref{sec:beta_coefficients} for details.
%Note that in the case that added single generation of vector-like fermions, that is, added $\mathbf{\bar{5}}$ and $\mathbf{10}$ representations for the vector-like fermions, the proton lifetime estimated in these mass ranges is not within the expected limit for proton lifetime by the Hyper-Kamiokande experiment, which means that the proton lifetime is inconsistent with the current experimental result or exceed that expected by the Hyper-Kamiokande experiment.

\begin{figure}[htbp]
\begin{minipage}[b]{0.5\linewidth}
    \centering
    \includegraphics[keepaspectratio,scale=0.65]{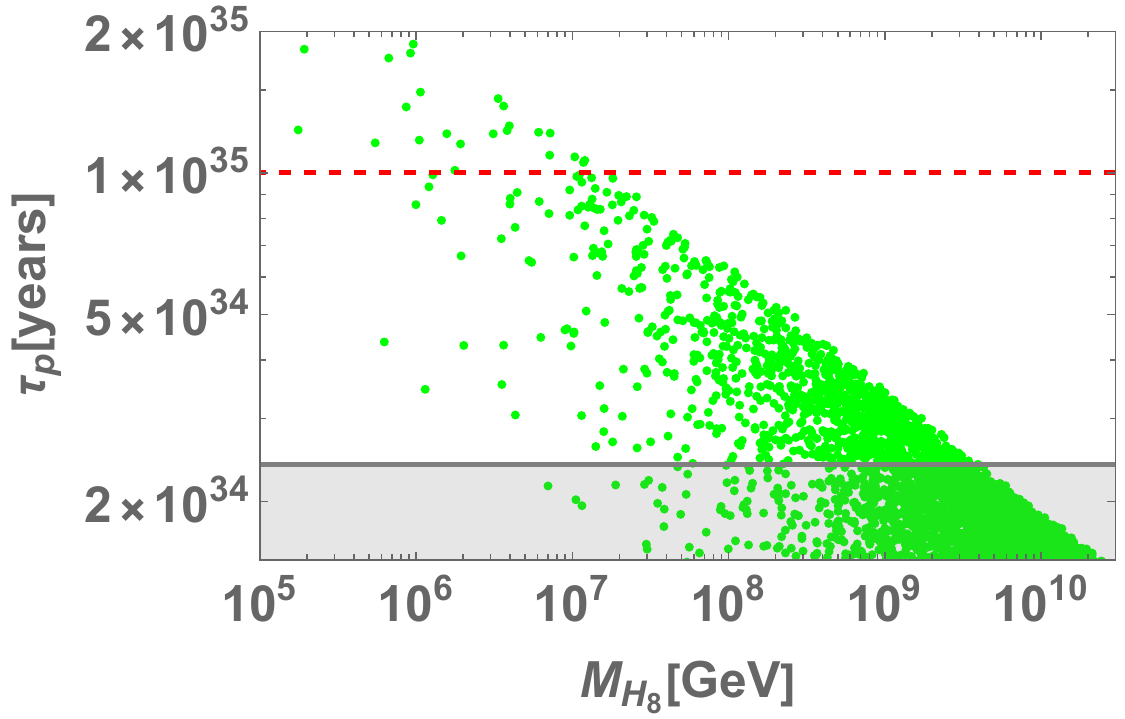}
    \subcaption{}
    \label{fig:M8tau}
\end{minipage}  
\begin{minipage}[b]{0.5\linewidth}
    \centering
    \includegraphics[keepaspectratio,scale=0.65]{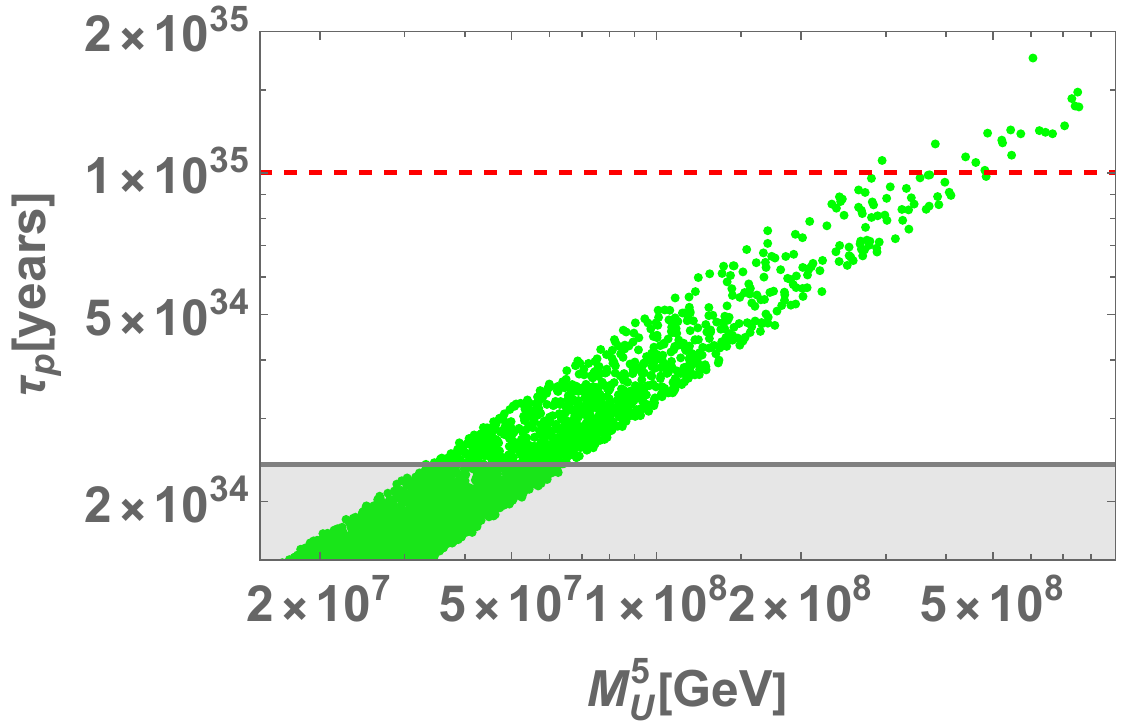}
    \subcaption{}
    \label{fig:MUtau}
\end{minipage}  
\caption{
The green points denote the range for \eqref{fig:M8tau}~the color octet scalar mass $M_{H_8}$, \eqref{fig:MUtau}~the 5th vector-like right-handed up quark mass and expected proton lifetime to achieve the unification of the SM gauge couplings with an accuracy of 1\% or less.
The excluded region for proton lifetime by Super-Kamiokande experiment is the gray shaded one as $\tau_p(p\to\pi^0{e^+})>2.4\times10^{34}$~years~\cite{Super-Kamiokande:2020wjk} and the red dashed line depicts the expected limit for proton lifetime by the Hyper-Kamiokande experiment, $\tau_p(p\to\pi^0{e^+})<1.0\times10^{35}$~years~\cite{Dealtry:2019ldr}.
}
\label{fig:M8MUtau}
\end{figure}

In Fig.~\ref{fig:M8MUtau}, the green points denote the allowed ranges for \eqref{fig:M8tau} the color octet scalar mass $M_{H_8}$, \eqref{fig:MUtau} the 5th vector-like right-handed up quark mass; plotted against the expected proton lifetime.
In that allowed range, the SM gauge couplings can unify successfully within an accuracy of 1\% or less.
We define the accuracy of the unification as a percentage difference between the energy scale of unified the SM gauge couplings $\alpha_1,~\alpha_2$ and the unified $\alpha_2,~\alpha_3$ couplings, where we express the SM gauge couplings for $\mathrm{U(1)_Y}$, $\mathrm{SU(2)_L}$, $\mathrm{SU(3)_C}$ as $\alpha_{i}\equiv{g_i}/4\pi$~$(i=1$-$3)$, respectively.
The excluded region for proton lifetime by the Super-Kamiokande experiment is the gray shaded region, $\tau_p(p\to\pi^0{e^+})>2.4\times10^{34}$~years~\cite{Super-Kamiokande:2020wjk}.
The red dashed line depicts the expected limit for proton lifetime by the Hyper-Kamiokande experiment, $\tau_p(p\to\pi^0{e^+})<1.0\times10^{35}$~years~\cite{Dealtry:2019ldr}.
Results of our analysis are the color octet scalar and the 5th vector-like right-handed up quark have masses in an intermediate scale.
Additionally, the 5th vector-like quark doublet and right-handed electron are required to have $\mathcal{O}(M_{\mathrm{GUT}})$ masses to unify the SM gauge couplings successfully.

\begin{figure}[htbp]
\begin{minipage}[b]{0.5\linewidth}
    \centering
    \includegraphics[keepaspectratio,scale=0.65]{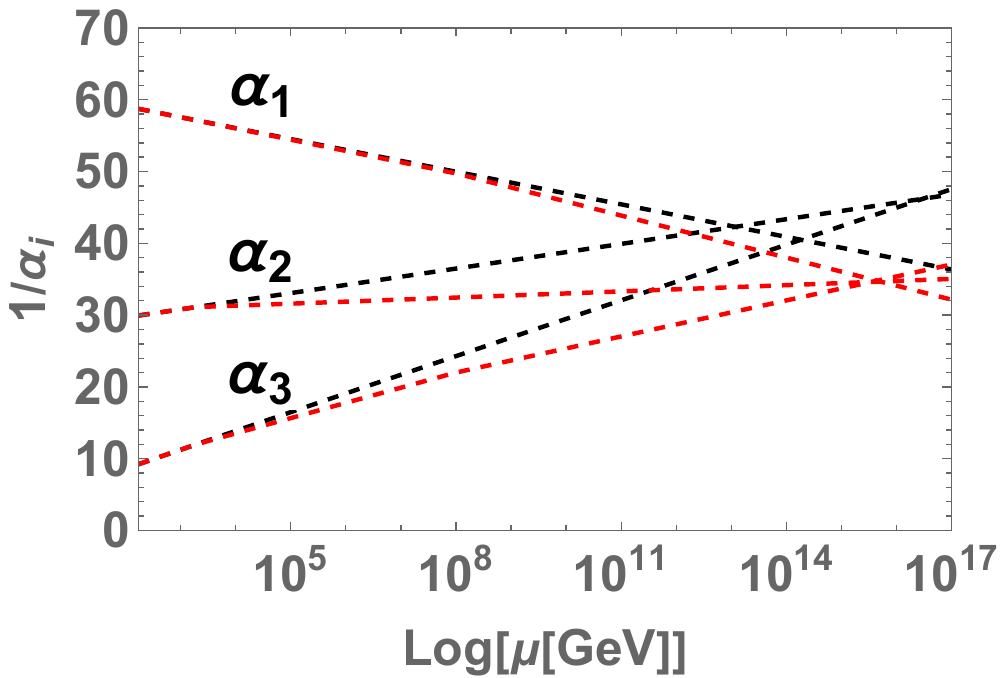}
    \subcaption{}
    \label{fig:unification}
\end{minipage}  
\begin{minipage}[b]{0.5\linewidth}
    \centering
    \includegraphics[keepaspectratio,scale=0.65]{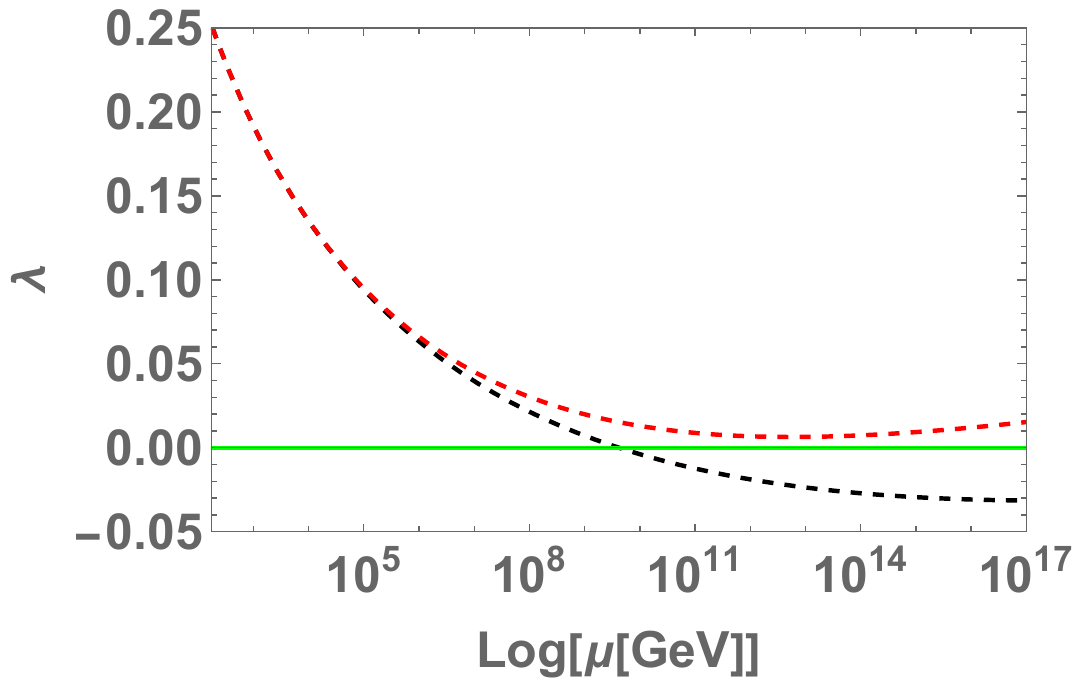}
    \subcaption{}
    \label{fig:lambda}
\end{minipage}  
\caption{The plot shows our results of running the SM couplings by solving the RGE for the fixed masses such that the 4th vector-like quark doublet has mass $M^4_Q=2000$~GeV, the real triplet scalar has mass $M_T=1500$~GeV, the 5th vector-like right-handed up quark and the color octet scalar have mass $M^5_U=M_{H_8}={10}^{8}$~GeV, and the other particles have the GUT scale masses, respectively.
In both figures, the red (black) dashed line shows the running of the SM couplings with (without) the contribution of the new particles.
In Fig.~\eqref{fig:unification}, the SM gauge couplings unify successfully at $M_{\mathrm{GUT}}\thickapprox5.1\times10^{15}$~GeV and the value for the unified gauge couplings is $\alpha_{\mathrm{GUT}}=\alpha_1=\alpha_2=\alpha_3\thickapprox1/34.7$.
In Fig.~\eqref{fig:lambda}, the SM Higgs quartic coupling is the positive value at all energy scales, and hence the SM Higgs potential is stabilized.
}
\label{fig:SMcoupling}
\end{figure}
To demonstrate gauge unification, we fix the masses such that the 4th vector-like quark doublet has mass $M^4_Q=2000$~GeV, the real triplet scalar has mass $M_T=1500$~GeV, the 5th vector-like right-handed up quark and the color octet scalar have mass $M^5_U=M_{H_8}={10}^{8}$~GeV, and the other particles have the GUT scale masses.

The left panel of Fig.~\ref{fig:SMcoupling} demonstrates the running of the SM gauge couplings by solving the RGEs.
The running of the SM gauge couplings for only the SM particles is depicted by the black dashed line.
The red dashed line is shown the running of the SM gauge couplings including the contribution of the new particles.
In the case included the new contribution, the SM gauge couplings unify successfully at $M_{\mathrm{GUT}}\thickapprox5.1\times10^{15}$~GeV and the value for the unified gauge couplings is $\alpha_{\mathrm{GUT}}=\alpha_1=\alpha_2=\alpha_3\thickapprox1/34.7$.
The expected proton lifetime for its decay mediated by the SU(5) gauge bosons is $\tau_p(p\to\pi^0{e^+})\approx{4.12\times10^{34}}~\mathrm{years}$.
This is consistent with the current experimental result for the proton lifetime given by the Super-Kamiokande experiment as $\tau_p(p\to\pi^0{e^+})>2.4\times10^{34}$~years~\cite{Super-Kamiokande:2020wjk}.
In addition, it is within the expected limit for proton lifetime by the Hyper-Kamiokande experiment, $\tau_p(p\to\pi^0{e^+})<1.0\times10^{35}$~years~\cite{Dealtry:2019ldr}.
Therefore, it is testable by the Hyper-Kamiokande experiment.

The right panel of Fig.~\ref{fig:SMcoupling} illustrates the RGE running of the SM Higgs quartic coupling $\lambda$.
%As is the case in the analysis about the running of the SM gauge couplings, we solve the RGE at 2-loop level.
The running of the SM Higgs quartic coupling for only the SM particles is depicted by the black dashed line.
The red dashed line is shown the running with the contributions of the new particles.
In addition, the horizontal green line depicts $\lambda=0$.
With the contributions from the new particles, the SM Higgs quartic coupling is positive for all energy scales.
Hence, the SM Higgs potential is stabilized.

%------------------------------------------------------------------------------%
%------------------------Summary and Discussions--------------------------------%
%------------------------------------------------------------------------------%
\section{Summary}
\label{sec:Summary}

We have proposed an SU(5) GUT model extended with two additional pairs of $\mathbf{10}$ representation vector-like fermions.
The VEV of a real $\mathrm{SU(2)_L}$ triplet coming from the $\mathbf{24}$ representation Higgs is used to explain the $W$ boson mass anomaly reported by the CDF collaboration.
We decomposed the vector-like fermions partly into vector-like quark doublets.
Those vector-like quark doublets acquire the mass from two sources;
through the Yukawa interaction of a real triplet via the type-$\rm{I\hspace{-.01em}I}$ seesaw-like mechanism and from the $\mathbf{24}$ representation Higgs.

We assumed that the mass for the vector-like quark doublet is expressed in terms of a real triplet mass.
Because the real triplet and heavy Higgs boson masses are almost the same, we could get constraints on the vector-like quark doublet mass from constraints of the heavy Higgs bosons.
By combining the constraints on the vector-like quark masses with those on the heavy Higgs boson masses, we could obtain the narrow allowed mass ranges for the vector-like quark doublet and the real triplet.
Assuming the Yukawa coupling for the 4th vector-like quark doublet is exactly one, the allowed mass ranges for the 4th vector-like quark doublet and the real triplet scalar are given by $1660<M^4_Q<2428$~GeV and $1400<M_T<1693$~GeV~(the heavy neutral Higgs boson limit) or $1660<M^4_Q<4759$~GeV and $1000<M_T<1693$~GeV~(the charged Higgs boson limit).
Therefore, our model can be tested by the search of these particles in the near future.

We set benchmark values for the mass eigenvalues of the relevant particles to solve the SM RGEs with contributions from new particles.
Thanks to the new contributions, the SM gauge couplings unify successfully at $M_{\mathrm{GUT}}\thickapprox5.1\times10^{15}$~GeV.
Also, the value for the unified gauge couplings is $\alpha_{\mathrm{GUT}}=\alpha_1=\alpha_2=\alpha_3\thickapprox1/34.7$.
In addition, the SM Higgs quartic coupling is positive for all energy scales, and thus the SM Higgs potential is stabilized.
Our model predicts the proton lifetime for decays mediated by the SU(5) gauge bosons as $\tau_p(p\to\pi^0{e^+})\approx{4.12\times10^{34}}$~years.
This is testable by future proton decay searches, for example the Hyper-Kamiokande experiment is expected to reach $\tau_p(p\to\pi^0{e^+})<1.0\times10^{35}$~years.

%-------- acknowledgement -------%
\vspace{1cm}
\noindent
{\large \bf Acknowledgement}
\vspace{1mm}

We would like to thank N. J. Benoit for reading our manuscript. 

%\newpage 
%---------------------------------------------------------%
%--------------- Appendix --------------------------------%
%---------------------------------------------------------%
\appendix
\section*{Appendix}
%\section{The minimal SU(5) Model}
\section{The RGE and beta coefficients}
\label{sec:beta_coefficients}
In our analysis of the running for the SM gauge couplings, we use the following RGE formulae,
\begin{equation}
    \mu\frac{dg_i}{d\mu}=g_i^3\Big[\beta_{g_i}(\mathrm{SM})+\beta_{g_i}(\mathrm{NEW})\Big],
\end{equation}
where $g_i$ $(i=1\mathrm{-}3)$ are the SM gauge couplings, $\beta_{gi}(\mathrm{SM})$ are the contributions of the SM particles, and $\beta_{gi}(\mathrm{NEW})$ are the new particle contributions, respectively.
We consider the contributions of the SM particles at the 2-loop level~\cite{Machacek:1983tz, Machacek:1983fi, Machacek:1984zw} and the new particle contributions at the 1-loop level;
\begin{equation}
    \beta_{g_i}(\mathrm{NEW})=\frac{1}{16\pi^2}\Big[b_i\times\theta(\mu-M)\Big],
\end{equation}
where $M$ is the mass of each field and the beta coefficients $b_i$ for each field are listed in Table~\ref{tab:betacoefficients}.
Here, $\theta(\mu-M)$ is a step function, which we use to add the new particle contributions for each particle.
The 1-loop beta coefficients for the SM particles are $(b_3^{\mathrm{SM}},b_2^{\mathrm{SM}},b_1^{\mathrm{SM}})=(-7,-19/6,41/10)$.
\begin{table}[h]
    \centering
    \begin{tabular}{|c|c|c|c|}
    \hline
         Fields&$b_3$&$b_2$&$b_1$\\
         \hline\hline
         $Q_{L,R}$&2/3&1&1/15\\
         \hline
         $U_{L,R}$&1/3&0&8/15\\
         \hline
         $E_{L,R}$&0&0&2/5\\
         \hline
         $T$&0&1/3&0\\
         \hline
         $H_8$&1/2&0&0\\
         \hline
    \end{tabular}
    \caption{The beta coefficients for each field.}
    \label{tab:betacoefficients}
\end{table}
\\
The 2-loop contributions of the SM particles are given by
\begin{align}
    \beta_{g_1}^{\mathrm{2-loop}}&=\left(\frac{1}{16\pi^2}\right)^2\left(\frac{199}{50}g_1^2+\frac{27}{10}{g^2_2}+\frac{44}{5}{g^2_3}-\frac{17}{10}y^2_t\right),\\
    \beta_{g_2}^{\mathrm{2-loop}}&=\left(\frac{1}{16\pi^2}\right)^2\left(\frac{9}{10}g_1^2+\frac{35}{6}{g^2_2}+12{g^2_3}-\frac{3}{2}y^2_t\right),\\
    \beta_{g_3}^{\mathrm{2-loop}}&=\left(\frac{1}{16\pi^2}\right)^2\left(\frac{11}{10}g_1^2+\frac{9}{2}{g^2_2}-26{g^2_3}-2y^2_t\right).
\end{align}
The RGE formula of the SM Higgs quartic coupling is given by
\begin{equation}
    \mu\frac{d\lambda}{d\mu}=\lambda\Big[\beta_{\lambda}^{\mathrm{1-loop}}+\beta_{\lambda}^{\mathrm{2-loop}}\Big],
\end{equation}
where
\begin{equation*}
    \begin{aligned}
        \beta_{\lambda}^{\mathrm{1-loop}}
        =&\frac{1}{16\pi^2}\Bigg[\frac{9}{4}\bigg(\frac{3}{25}g^4_1+\frac{2}{5}g_1^2{g_2^2}+g_2^4\bigg)-12y_t^4-\bigg(\frac{9}{5}g_1^2+9g_2^2\bigg)\lambda+12y_t^2{\lambda}+12\lambda^2\Bigg],\\
        \beta_{\lambda}^{\mathrm{2-loop}}
        =&\left(\frac{1}{16\pi^2}\right)^2\Bigg[-\frac{3411}{1000}{g_1^6}-\frac{1677}{200}{g_1^4{g_2^2}}-\frac{289}{40}{g_1^2{g_2^4}}+\frac{305}{8}{g_2^6}-\frac{9}{2}{g_2^4}{y_t^2}\\
        &+\frac{3}{5}{g_1^2}\bigg(-\frac{57}{10}{g_1^2}+21{g_2^2}\bigg){y_t^2}-\frac{16}{5}{g_1^2}{y_t^4}-64{g_3^2}{y_t^4}+60{y_t^6}\\
        &-\bigg(\frac{1119}{200}{g_1^4}+\frac{117}{20}{g_1^2}{g_2^2}+\frac{73}{8}{g_2^4}\bigg)+10\bigg(\frac{17}{20}{g_1^2}+\frac{9}{4}{g_2^2}+8{g_3^2}\bigg)\lambda{y_t^2}\\
        &-3{y_t^4}{\lambda}+18\bigg(\frac{3}{5}{g_1^2}+3{g_2^2}\bigg)\lambda^2-72{y_t^2}{\lambda^2}-78\lambda^3\Bigg].
    \end{aligned}
\end{equation*}
%------------------------------------------------------------------------------%
%--------------------------    References    --------------------------------------%
%------------------------------------------------------------------------------%
%\newpage

\end{document}